\documentclass[a4paper]{jpconf}
\usepackage{graphicx}
\usepackage{amsfonts,verbatim}
\usepackage[T1]{fontenc}

\def\Journal#1#2#3#4{#4 \emph{#1} {\bf #2} #3}

\begin{document}
\title{A new special class of Petrov type D vacuum space-times in dimension
five.}

\author{Alfonso Garc\'{\i}a-Parrado G\'omez-Lobo $^1$ and Lode Wylleman $^{2,3}$}

\address{$^1$ Centro de Matem\'atica, Universidade do Minho.
4710-057 Braga, Portugal}
\address{$^2$ Utrecht University, Department of Mathematics, Budapestlaan 6, 3584 CD Utrecht, the Netherlands}
\address{$^3$ Ghent University, Department of Mathematical Analysis IW16, Galglaan 2, 9000 Ghent, Belgium}
\ead{alfonso@math.uminho.pt}
\ead{lwyllema@ugent.be}

\begin{abstract}
Using extensions of the Newman-Penrose and  Geroch-Held-Penrose
formalisms to five dimensions, we invariantly classify all Petrov
type $D$ vacuum solutions for which the Riemann tensor is isotropic
in a plane orthogonal to a pair of Weyl alligned null
directions \footnote{We dedicate this work to the memory of
Professor S. B. Edgar (1945-2010).}.
\end{abstract}

\section{The role of the Newman-Penrose (NP) and Geroch-Held-Penrose (GHP) formalisms in General Relativity}
The NP formalism~\cite{NEWMAN-PENROSE} consists in an explicit
expansion of the Ricci and Bianchi identities in a {\em null
tetrad}, yielding a set of equations (henceforth called NP
equations) for the {\em spin coefficients} and {\em curvature
scalars}. Under certain restrictions (vacuum, algebraic type of the
Weyl and/or Ricci tensors, etc) one can choose a null tetrad in which
some curvature scalars vanish bringing a simplification to the NP
equations. In some cases it is possible to perform a complete study
of the consistency conditions of the constraints and their higher
order differentials, thus enabling the classification of the family
of solutions abiding by the restrictions imposed.
The GHP formalism~\cite{GHP} further exploits the invariance of the
null tetrad directions under the group of boost and spin
transformations in order to work with those variables which are {\em
covariant} under this group. This reduces the number of variables
and simplifies the equations. From these considerations, we deduce
that the GHP formalism is best suited for situations in which the
boost-spin group is the natural invariance group in the
geometric set-up. An important example of this can be found in the
case of Petrov type D solutions, where the two principal null
directions of the Weyl tensor are naturally invariant under boosts,
and their orthogonal 2-plane is invariant under spins (see
\cite{EDGAR-D,VICKERS-D} for further details).

\section{A generalisation of the NP and GHP formalisms to dimension five}
The techniques described in the previous paragraph have been developed
for the case of a four-dimensional (4D) space-time, but the ideas
are general for any dimension as explained in
\cite{PRAVDA-BIANCHI,PRAVDA-RICCI,PRAVDA-GHP}. In our work we
analyse these ideas for the particular case of a 5D space-time. In
this particular case, one can still rely on spinor calculus to
introduce the essential quantities of the formalisms, in the same
way as in the 4D case. The basics of spinor calculus in a 5D
space-time can be found in \cite{SPIN-5D} and we stress that the
rationale behind many of our definitions lies in the use of spinors,
even though this will not be made explicit in the present work.

We start by introducing a {\em semi-null pentad} defined as follows
\begin{equation}
N\equiv\{l^a, n^a, m^a,\bar{m}^a,u^a\},\;\quad
l^an_a=-1\;,\quad m^a\bar{m}_a=1\;,\quad u^a u_a=1.
\label{eq:semi-null-pentad}
\end{equation}
The {\em frame derivations} associated to each element of this frame are denoted as follows:
\begin{equation}
D\equiv l^a\nabla_a\;,\quad
\Delta\equiv n^a\nabla_a\;,\quad
\delta\equiv m^a\nabla_a\;,\quad
\bar\delta\equiv\bar m^a\nabla_a\;,
\quad\mathcal{D}\equiv u^a\nabla_a.
\label{eq:frame-derivations}
\end{equation}
Next we need to define the {\em spin coefficients} and the {\em
curvature scalars}. Some of them coincide with the quantities
appearing in the standard 4D NP formalism (this is to be expected as
we work in a frame obtained from the NP null tetrad by just adding
the element $u^a$). Here we just enumerate the quantities.
\begin{itemize}

\item
Twelve NP 4D spin coefficients: $\alpha$, $\beta$, $\gamma$,
$\epsilon$, $\kappa$, $\lambda$, $\mu$, $\nu$, $\pi$, $\rho$,
$\sigma$, $\tau$.

\item
Ten complex 5D spin coefficients: $\zeta$, $\eta$, $\theta$,
$\chi$, $\omega$, $\phi$, $\xi$, $\upsilon$, $\psi$, $\varsigma$.

\item
Six real 5D spin coefficients: $\mathfrak{a}$, $\mathfrak{b}$,
$\mathfrak{c}$, $\mathfrak{d}$, $\mathfrak{e}$, $\mathfrak{f}$.
$$
2\times 12+2\times 10+6=50\ \mbox{real Ricci rotation coefficients}.
$$

\item
Five complex 4D Weyl scalars: $\Psi_0$, $\Psi_1$, $\Psi_2$,
$\Psi_3$, $\Psi_4$,

\item
Eleven complex 5D Weyl scalars: $\ ^*\Psi_0$,\ $\ ^*\Psi_1$,\
$\Psi^*_1$,\ $\ ^*\Psi_2$,\ \ $\Psi^*_2$,\ $\ ^*\Psi_3$,\ \
$\Psi^*_3$,\ $\Psi^*_4$,\ $\Psi_{01}$, $\Psi_{02}$, $\Psi_{12}$.

\item
Three real 5D Weyl scalars: $\Psi_{00}$, $\Psi_{11}$, $\Psi_{22}$.

\item
4D NP Ricci scalars: Real: $\Phi_{00}$, $\Phi_{11}$, $\Phi_{22}$,
Complex: $\Phi_{01}$, $\Phi_{02}$, $\Phi_{12}$.

\item
5D trace-free Ricci scalars: Real: $\Omega$, $\ ^*\Phi_{01}$, $\
^*\Phi_{12}$, Complex: $\ ^*\Phi_{02}$.

\item
Scalar curvature: $\Lambda$.
$$
(2\times16+3)+(2\times 4 +6)+1=50\ \mbox{independent real Riemann
tensor components.}
$$
\end{itemize}

The procedure is now to expand the Ricci and Bianchi identities in
the semi-null pentad (\ref{eq:semi-null-pentad}) and write them out
in terms of the scalar quantities just introduced. The resulting
equations are referred to as the 5D NP equations and they can be
regarded as an {\em extension} of the well-known 4D NP equations.
One can also compute the commutation relations of the operators
defined in (\ref{eq:frame-derivations}) in terms of the spin
coefficients and again a set of equations is obtained which extends
to dimension five the known commutation relations of the NP
formalism.

Under the action of the boost-spin group the semi-null pentad
transforms according to
\begin{eqnarray}
&& l^a\mapsto z\bar z\, l^a\;,\quad n^a\mapsto \frac{n^a}{z \bar
z}\;,\quad m^a\mapsto \frac{z}{\bar z}\,m^a\;,\quad{\bar m}^a\mapsto
\frac{\bar z}{z}\,{\bar m}^a\;,\quad u^a\mapsto u^a\label{eq:spin}
\end{eqnarray}
for some complex number $z\in\mathbb{C}$. A scalar quantity $Q$ is
said to be {\em weighted} if under (\ref{eq:spin}) it transforms as
\begin{equation}
Q\mapsto z^p\bar z^q Q,
\label{eq:scalar-weight}
\end{equation}
for some integers $p$ and $q$ (the pair $(p,q)$ being called the weight of $Q$). It is not difficult to check that all the curvature
scalars are weighted quantities whereas only a subset of the spin
coefficients is weighted. Now, we can follow a procedure similar to
the one presented in \cite{GHP} and introduce new derivations {\em
covariant} under boosts and spins. These new derivations turn out to
be uniquely defined and are given by~\cite{SPIN-5D}
\begin{eqnarray}
&&\mbox{\th}Q\equiv(D-p\epsilon-q\bar\epsilon)Q\;,\quad
\mbox{\th}'Q\equiv(\Delta-p\gamma-q\bar\gamma)Q\;,\quad
\nonumber\\
&&
\quad\mbox{\dh}Q\equiv(\delta-p\beta-q\bar\alpha)Q\;,\quad\mbox{\dh}'Q\equiv(\bar\delta-p\alpha-q\bar\beta)Q\;,\quad
\widehat{\mathcal{D}}Q\equiv(\mathcal{D}-p\theta-q\bar\theta)Q\;,
\end{eqnarray}
where $Q$ is a quantity with weight $(p,q)$ as in
(\ref{eq:scalar-weight}). They are
referred to as {\em GHP operators}. Regarding the covariance under a boost-spin transformation (\ref{eq:spin}) one specifically has
\begin{eqnarray}
&&\mbox{\th}Q\mapsto z^{1+p}\bar{z}^{1+q}\mbox{\th}Q\;,\quad
\mbox{\th}'Q\mapsto z^{p-1}\bar{z}^{q-1}\mbox{\th}Q\;,
\nonumber\\
&&\mbox{\dh}Q\mapsto z^{p-1}\bar{z}^{q+1}\mbox{\dh}Q\;,\quad
\mbox{\dh}'Q\mapsto
z^{p+1}\bar{z}^{q-1}\mbox{\dh}'Q\;,\quad\widehat{\mathcal{D}}Q\mapsto
\widehat{\mathcal{D}}Q\;.
\end{eqnarray}
One can now extract the part of the NP equations which only involves
derivatives of weighted quantities and obtain a set of equations
which we call the 5D GHP equations. The remaining NP equations are
absorbed in the commutator relations of the GHP operators.

\section{The ${\mathcal A}$-class and its invariant classification}
The {\em null alignment theory}
\cite{ALIGNMENT} provides an algebraic
classification of any tensor in a Lorentzian vector space of
arbitrary dimension. For the Weyl tensor this implies a generalization of the 4D {\em Petrov types}. These are characterised by the {\em
Weyl aligned null directions} (WANDS) and their {\em alignment
order}. The WANDs generalise the notion of {\em principal null
direction} of a 4D Weyl tensor. An important difference lies in the
fact that the number of WANDs might be infinite or zero (the latter
being in fact the generic case) in higher than four dimensions,
whereas it is exactly four (counting multiplicity) in the 4D case. We refer the reader to
\cite{ALIGNMENT} for further details.

In view of the complete classification of the 4D Petrov type D
vacuum space-times (containing the well-known vacuum black hole
solutions)~\cite{Kinnersley,Debeveretal2,Garcia,SKMHH,EDGAR-D}
it is natural to endeavor the classification of 5D Petrov type D
vacua (or {\em Einstein spaces}). By definition, these are space-times (possibly admitting a
cosmological constant $\Lambda$) with vanishing trace-free Ricci
tensor and exhibiting a pair of double WANDs spanned by $l^a$ and
$n^a$ at each point. Taking these as the first two vectors of a
semi-null pentad, the only surviving components of the Weyl tensor
are the zero boost-weight Weyl scalars
\begin{equation}\label{Psis}
\Psi_{11}\;,\quad \Psi_{2}\;,\quad \Psi_{02}\;,\quad
\Psi^*_2\;,\quad \ ^*\Psi_2.
\end{equation}
It turns out that the resulting GHP equations and its consequences
are rather unmanageable, although some general properties (even in general dimensions) were deduced in \cite{PravdaOrtaggio,DurkeeReall}.
In order to simplify the computations
further, we restricted ourselves to the class $\mathcal A$, defined
by the property that the Riemann (or Weyl) tensor is isotropic in a
plane orthogonal to $l^a$ and $n^a$. Taking the pentad vectors $m^a$
and ${\bar m}^a$ along the complex null directions of this plane and
$u^a$ along its normal, only the $(0,0)$-weighted Weyl scalars
$\Psi_{11}$ and $\Psi_{2}$ are possibly non-zero.

The procedure followed consists in imposing the conditions just
described on the extended GHP equations and analyse the
corresponding consistency conditions. The details of the
integrability analysis will be presented in \cite{ParradoWylleman}.
Here the possible cases are summarised in table \ref{a:class}.  Each
case is characterised by algebraic relations fulfilled by the
quantities $\Psi_{11}$ and $\Psi_2$. The case $\Psi_{11}=0$ is a
direct generalization of the 4D solutions, and can be fully integrated, just as the family characterized by the condition $h_1$. The case
$\Psi_2=2\Psi_{11}$ was completely integrated in
\cite{DurkeeReall} and only here the pair of double WANDs is not
unique. The generic metric corresponding to
$\Psi_2=-2\Psi_{11}$ has trivial isometry group and admits free functions, whereas those for all
other cases only depend on a number of invariantly defined continuous parameters (`global charges'). It is possible to obtain
invariant information (without performing the
actual integration of the solutions)
in the same spirit as
e.g.~\cite{EDGAR-D}. The Karlhede bound refers to the number of covariant derivatives of the Riemann tensor needed in the invariant classification algorithm~\cite{SKMHH}. The symbol $s$ ($r$) denotes the dimension of the isotropy (isometry) group.

\begin{table}
\caption{\label{a:class} Characterizing relations for the
subclasses of $\mathcal A$. The symbol $h_1$ denotes the condition
$\Psi_2=\bar\Psi_2$, $\Psi_2\bar\Psi_2\neq 4\Psi^2_{11}\neq 0$ and
the symbol $h_2$ stands for the condition
$\Psi_2\bar\Psi_2=4\Psi^2_{11}$, $\bar\Psi_2\neq\Psi_2$.}
\vspace{.2cm}
\begin{center}
\begin{tabular}{cccccc}
 \br
Subclass & $\Psi_{11}=0$ & $h_1$& $h_2$&$\Psi_2=2\Psi_{11}$&$\Psi_2=-2\Psi_{11}$\\
\mr
Karlhede bound  & $\leq 2$    &$1$  & $\geq 1$ &$\leq 2$&$\geq 1$\\
Global charges &$\leq 4$  & 2 & $\leq 3$&$\leq 1$& Free functions\\
$s$&$\leq 2$&2&$\leq 2$&3 or 4&$\leq 2$\\
$r$&$\geq 2$&6&$\geq 3$&7 or 9&$\geq 0$\\
\br
\end{tabular}
\end{center}
\end{table}

\section{Conclusion and further research}
We performed the invariant classification of a particular class $\cal A$
of five-dimensional vacuum metrics of Weyl-Petrov type D,
possibly admitting a cosmological constant. Some issues have been
left open in this work. Perhaps the most important one is a
complete integration of the $h_2$ and $\Psi_2=-2\Psi_{11}$ cases in table \ref{a:class}, which would enable us to carry out a detailed study of the
properties of these solutions. This would require the
further development of integration techniques for the differential
equations arising in the 5D GHP equations, in the line of those
largely explored by Brian Edgar in the 4D case (see e.g.\ \cite{EL,ER,EDGAR-D}). Another interesting avenue is to
explore other Petrov type D subtypes different to the
$\mathcal{A}$-class (the generalisation of our analysis to obtain the
generic five-dimensional type D vacuum solution seems to be rather
involved).

\section*{Acknowledgments}
AGP is supported by the Research Centre of
Mathematics of the University of Minho (Portugal) through the ``Funda\c{c}\~ao para a Ci\^encia e a Tecnologia'' (FCT) Pluriannual
Funding Program. LW is supported by a BOF Research Grant (UGent) and a FWO mobility grant.

\end{document}